\documentclass[reprint,amsmath,amssymb,aps,superscriptaddress]{revtex4-1}
\usepackage{graphicx}
\usepackage{sidecap}
\usepackage{array}
\usepackage{subfigure}
\usepackage{color}
\usepackage{amsmath}
\usepackage{booktabs}

\begin{document}

\title{A review of elliptic flow of light nuclei in heavy-ion collisions at RHIC and LHC energies}

\author{Md. Rihan Haque}
\email{r.haque@uu.nl}
\affiliation{Utrecht University, P.O. Box 80000, 3508 TA Utrecht, The Netherlands}
\author{Chitrasen Jena}
\email{cjena@iisertirupati.ac.in}
\affiliation{School of Physical Sciences, National Institute of
  Science Education and Research, Jatni 752050, India}
\affiliation{Indian Institute of Science Education and
  Research, Tirupati 517507, India}
\author{Bedangadas Mohanty}
\email{bedanga@niser.ac.in}
\affiliation{School of Physical Sciences, National Institute of
  Science Education and Research, Jatni 752050, India}

\begin{abstract}
 We present a review of the measurements of elliptic flow ($v_{2}$) of
 light nuclei ($d$, $\overline{d}$, $t$, $^{3}He$ and $^{3}\overline{He}$) 
 from the RHIC and LHC experiments. Light (anti-)nuclei $v_{2}$ have
 been compared with that of \mbox{(anti-)proton}. We observed similar trend
 in light nuclei $v_{2}$ as in identified hadron $v_{2}$ with respect
 to the general observations such as $p_{T}$ dependence, low $p_{T}$
 mass ordering and centrality dependence. We also compared the
 difference of nuclei and anti-nuclei $v_{2}$ with the corresponding
 difference between $v_{2}$ of proton and anti-proton at various collision
 energies. Qualitatively they depict similar behavior.  We also
 compare the data on light nuclei $v_{2}$ to various theoretical
 models such as blast-wave and coalescence. We then present a
 prediction of $v_{2}$  for $^{3}He$ and $^{4}He$ using coalescence
 and blast-wave models.

\end{abstract}

 \maketitle

 \section{Introduction}

The main goals of high energy heavy-ion collision 
experiments have primarily been to study the properties
of Quark Gluon Plasma (QGP) and the other phase structures in the QCD
phase diagram~\cite{QCD_theo, whitepapers,QCD_theo2,QCD_expt3}. The
energy densities created in these high energy collisions are similar
to that found in the universe, microseconds after the Big
Bang~\cite{whitepapers, ALICEPerformace,Early_Univ}. Subsequently,
the universe cooled down to form nuclei. It is expected that
high energy heavy-ion collisions will allow to study the production of
light nuclei such as $d$, $t$, $^{3}He$ and their corresponding  
anti-nuclei. There are two possible production mechanisms  
for light (anti-)nuclei. The first mechanism is thermal production of 
nucleus-antinucleus pairs in elementary nucleon-nucleon 
or parton-parton interactions~\cite{Ratio_nucl2,Ratio_nucl,CFO_nucl}. 
However, due to their small ($\sim$few MeV) binding energies, the
directly produced nuclei or anti-nuclei are likely to break-up in the
medium before escaping. The second mechanism is via final state
coalescence of produced (anti-)nucleons or from transported  
nucleons ~\cite{nucl_prod1,nucl_prod2,nucl_prod3,nucl_prod4,
nucl_prod5,nucl_prod6,nucl_prod7,nucl_prod8}. 
The quark coalescence as a mechanism of hadron production at
intermediate transverse momentum has been well established by 
studying the number of constituent quark (NCQ) scaling for $v_{2}$ of 
identified hadrons measured at RHIC~\cite{part_coll1,part_coll2, 
part_coll3,v2_BES_PID,v2_200_QM,PID_v2_PHENIX,PID_pT_PHENIX}. 
Light nuclei may also be produced via coalescence of quarks similar to 
the hadrons. But the nuclei formed via quark coalescence is highly unlikely to 
survive in the high temperature environment due to their small binding
energies. In case of hadron 
formation by quark coalescence, the momentum space distribution of 
quarks are not directly measurable in experiments. However, in case of 
nucleon coalescence, momentum space distributions of both the constituents 
(nucleons) and the products (nuclei) are measurable in heavy-ion collision 
experiments. Therefore, measurements of $v_{2}$ of light nuclei provides a 
tool to understand the production mechanism of light nuclei and freeze-out 
properties at a later stage of the evolution. It also provides an excellent
opportunity to understand the mechanism of coalescence at work in high
energy heavy-ion collisions. 

The production of light (anti-)nuclei has been studied extensively at
lower energies in Bevelac at LBNL~\cite{bevalac1,bevalac2}, AGS at RHIC~\cite{ags1,ags2}
and SPS at CERN~\cite{sps1,sps2}. In the AGS experiments, it was found
that the coalescence parameter ($B_{2}$) is of similar magnitude for
both $d$ and $\overline{d}$ indicating similar freeze-out hypersurface
of nucleons and anti-nucleons. Furthermore, the dependence of $B_{2}$ 
on collision energy and $p_{T}$ indicated that light nuclei production
is strongly influenced by the source volume and transverse expansion 
profile of the system~\cite{sps2,polleri}.
In this paper, we review the results of elliptic flow of light nuclei
measured at RHIC and LHC and discuss the possible mechanisms
for the light nuclei production. 

The paper is organized as follows. Sec. II briefly describes the
definition of $v_{2}$, identification of light \mbox{(anti-)nuclei} in the
experiments and measurement of $v_{2}$ of light (anti-)nuclei. In Sec. III, we
present the $v_{2}$ results for minimum bias collisions from various
experiments. We also discuss the centrality dependence, difference
between nuclei and anti-nuclei $v_{2}$ as well as the energy
dependence of deuteron $v_{2}$. In the same section, we present the
atomic mass number scaling and also compare the experimental results
with various theoretical models. Finally in Sec. IV, we summarize our
observations and discuss the main conclusions of this review. 


\begin{table*}[t]
\small
\caption{Available measurements of light nuclei $v_{2}$}
{\renewcommand{\arraystretch}{1.3}
\begin{tabular} {l c l c l c l c l}
\hline 
\hline
  Experiment & Nuclei & \hspace{0.15cm} $\sqrt{s_{NN}}$ (GeV)  & Centrality   \\ [0.6ex]
 \hline
 STAR~\cite{v2_BES_nucl} & $\begin{matrix} d, \overline{d}, t, \\ ^{3}He, ^{3}\overline{He} \end{matrix}$
  & $\begin{matrix} 7.7, 11.5, 19.6, 27,\\ 39, 62.4, 200 \end{matrix}$ &
 $\begin{matrix} \text{0-80\%, 0-30\%, 30-80\%} \\ \text{(0-10\%, 10-40\%, 40-80\% in
   200 GeV)} \end{matrix}$  \\ [0.2ex] 
 
 PHENIX~\cite{nucl_PHENIX} & $d$+$\overline{d}$  & \hspace{0.4cm} 200 &0-20\%, 20-60\% \\   [0.2ex] 
 
 ALICE (Preliminary)~\cite{v2_nucl_ALICE}  &  $d$+$\overline{d}$  & \hspace{0.4cm} 2760 & 0-5\%, 5-10\%, 10-20\%, 20-30\%,
 30-40\%, 40-50\%  \\ [0.2ex] 
\hline
\hline
\end{tabular}
}
 \label{table:nuclei_data}
\end{table*}

\section{Experimental Method}

\subsection{Elliptic flow ($v_{2}$)} 
The azimuthal distribution of produced particles in heavy-ion
collision can be expressed in terms of a Fourier series,
 \begin{equation}
 \frac{dN}{d(\phi-\Psi_{r})} \propto 1+ \sum\limits_{n} 2{v_{n}}\cos[n(\phi-\Psi_{r})], 
 \label{eq:fourier}
 \end{equation}
\noindent where $\phi$ is the azimuthal angle of produced particle, $\Psi_{r}$
is called the reaction plane angle and the Fourier co-efficients
$v_{1}$, $v_{2}$ and so on are called flow co-efficients~\cite{voloshin1}. 
$\Psi_{r}$ is defined as the angle between the impact parameter vector 
and the x-axis of the reference detector in the laboratory frame. Since it
is impossible to measure the direction of impact parameter in
heavy-ion collisions, a proxy of $\Psi_{r}$ namely the event plane
angle $\Psi_{n}$ is used for the flow analysis in heavy-ion
collisions~\cite{art1}. The $v_{2}$ is measured with respect to the
2$^{nd}$ order event plane angle $\Psi_{2}$~\cite{art1}.  $\Psi_{2}$
is calculated using the azimuthal distribution of the produced
particles. In an event with $N$ particles, the event plane angle
$\Psi_{2}$ is defined as~\cite{art1}:   
  \begin{equation}
  \Psi_{2}\ = \frac{1}{2}\tan^{-1}(\frac{Y_{2}}{X_{2}}) .
  \label{eq:psi2}
 \end{equation}
 $X_{2}$ and $Y_{2}$ are defined as
 \begin{subequations}
 \label{sub:QxQy}
  \begin{align}
 X_{2} = \sum\limits_{i=1}^{N} w_{i}\cos(2\phi_{i}) , \\
 Y_{2} = \sum\limits_{i=1}^{N} w_{i}\sin(2\phi_{i})  ,
  \end{align}
 \end{subequations}
\noindent where $w_{i}$ are weights given to each particle to 
optimise the event plane resolution~\cite{art1,ptweight}.
Usually the magnitude of particle transverse momentum $p_{T}$ is used 
as weights as the $v_{2}$ increases with $p_{T}$. Special techniques
are followed while calculating the event plane angle so that it does 
not contain the particle of interest whose $v_{2}$ is to be calculated 
(self-correlation) and also the non-flow effects (e.g., jets and short 
range correlations) are removed as much as possible~\cite{art1, 
v2_BES_PID, v2_BES_nucl}. Heavy-ion experiments use 
the $\eta$-sub event plane method to calculate the elliptic 
flow of identified hadrons as well as for light nuclei. 
In this method, each event is divided into two sub-events in two different 
$\eta$-windows ($e.g.,$ positive and negative $\eta$). Then two sub-event 
plane angles are calculated with the particles in each sub-event. Each 
particle with a particular $\eta$ is then correlated with the sub event plane of
the opposite $\eta$. This ensures 
that the particle of interest is not included in the calculation of 
event plane angle. A finite $\eta$ gap is applied between the two 
sub-events to reduce short range correlations which does
not originate from flow. 

The distribution of the event plane angles 
should be isotropic in the laboratory frame for a
azimuthally isotropic detector. If the distribution of the event plane
angles is not flat in the laboratory frame (due to detector acceptance
and/or detector inefficiency) then special techniques are applied to
make the distribution uniform. The popular methods to make the
$\Psi_{2}$ distribution uniform is the $\phi$-weight and recentering
~\cite{AGS1,AGS2}. In the $\phi$-weight method, one takes the actual azimuthal
distribution of the produced particle, averaged over large sample of
events, and then uses inverse of this distribution as weights while
calculating the correlation of the particles with the event plane
angle~\cite{AGS1,AGS2}. In the recentering method, one
subtracts $\langle X_{n} \rangle$ and $\langle Y_{n} \rangle$ from the
event-by-event $X_{n}$ and $Y_{n}$, respectively, where $\langle X_{n}
\rangle$ and $\langle Y_{n} \rangle$ denotes the average of $X_{n}$
and $Y_{n}$ over a large sample of similar events. 
 The main disadvantage of applying one of these methods is that it
does not remove the contribution from higher flow harmonics. 
Therefore, another correction method known as the shift  
correction is used to remove the effects coming from higher flow harmonics. 
In this method, one fits the $\Psi_{2}$ distribution (after applying 
$\phi$-weight and/or recentering method) averaged over all events, with a
Fourier function. The Fourier co-efficients from this fit (obtained as
fit parameters) are used to shift the $\Psi_{2}$ of each event, to
make the distribution uniform in the laboratory frame~\cite{AGS1,AGS2}.  
 
Since the number of particles produced in heavy-ion collisions are
finite, the calculated event plane angle $\Psi_{2}$ may 
not coincide with $\Psi_{r}$. For this reason, the measured
$v^{obs}_{2}$ with respect to $\Psi_{2}$ is corrected with 
the event plane resolution factor $R_{2}$, where
\begin{equation}
 R_{2}\ =\ \langle \cos[2(\Psi_{2}-\Psi_{r})] \rangle.
\label{eq:reso2}
\end{equation}

In order to calculate the event plane resolution, one calculates two
sub-event plane angles $\Psi^{a}_{2}$ and $\Psi^{b}_{2}$, where $a$
and $b$ corresponds to two independent sub-events. If the
multiplicities of each sub-event are approximately half of the full
event plane, then the resolution of each of sub-event plane can be
calculated as~\cite{voloshin1,art1}, 
\begin{equation}
\langle \cos[2(\Psi^{a}_{2}-\Psi_{r})] \rangle\ =\
\sqrt{\langle \cos[2(\Psi^{a}_{2}-\Psi^{b}_{2})] \rangle}.
\label{eq:subreso}
\end{equation}
However, the full event plane resolution can  be expressed as,
\begin{eqnarray}
\langle \cos[2(\Psi_{2}-\Psi_{r})] \rangle\ =\
\frac{\sqrt{\pi} }{ 2\sqrt{2}} \chi_{2}\exp(-\chi^{2}_{2}/4) \nonumber \\ 
\times [I_{0}(\chi^{2}_{2}/4) + I_{2}(\chi^{2}_{2}/4)],
\label{eq:chi2find}
\end{eqnarray}
where, $\chi_{2}$ = $v_{2}/\sigma$ and $I_{0}$, $I_{2}$ are modified Bessel
functions~\cite{voloshin1,art1}. The parameter $\sigma$ is inversely
proportional to the square-root of $N$, the number of particles used
to determine the event plane~\cite{voloshin1,art1}. To calculate the
resolution for full event plane ($\Psi_{2}$), one has to solve the
Eq. (\ref{eq:chi2find}) iteratively for the value of $\chi_{2}$ using
the subevent plane resolution ($\langle \cos[2(\Psi^{a}_{2}-\Psi_{r})]
\rangle$) which is calculated experimentally using
Eq. (\ref{eq:subreso}). The  $\chi_{2}$ value  is then multiplied
with $\sqrt{2}$ as $\chi_{2}$ is propotional to $\sqrt{N}$, and
re-used in Eq. (\ref{eq:chi2find}) to calculate the resolution of
the full event plane.
In a case of very low magnitudes, the full event plane
resolution can be approximately given as~\cite{voloshin1,art1},
\begin{eqnarray}
\langle \cos[2(\Psi_{2}-\Psi_{r})] \rangle\ =\
\sqrt{2}\langle \cos[2(\Psi^{a}_{2}-\Psi_{r})] \rangle\ \nonumber \\
=\ \sqrt{2\langle \cos[2(\Psi^{a}_{2}-\Psi^{b}_{2})] \rangle}.
\label{eq:fullreso}
\end{eqnarray}
The procedure for calculating full and sub-event plane resolutions
using sub-event plane angles and various approximations are
discussed in detail in~\cite{voloshin1,art1}.

\subsection{Data on light nuclei} 
 For this review, we have collected light nuclei $v_{2}$ data from the
 STAR~\cite{v2_BES_nucl} and PHENIX~\cite{nucl_PHENIX} 
 experiments at RHIC and ALICE experiment at
 LHC~\cite{v2_nucl_ALICE}. The table~\ref{table:nuclei_data}
 summarises the measurement of light nuclei $v_{2}$ available till
 date.

 \begin{figure*}[t!]
 \centering
 \includegraphics[totalheight=7.64cm]{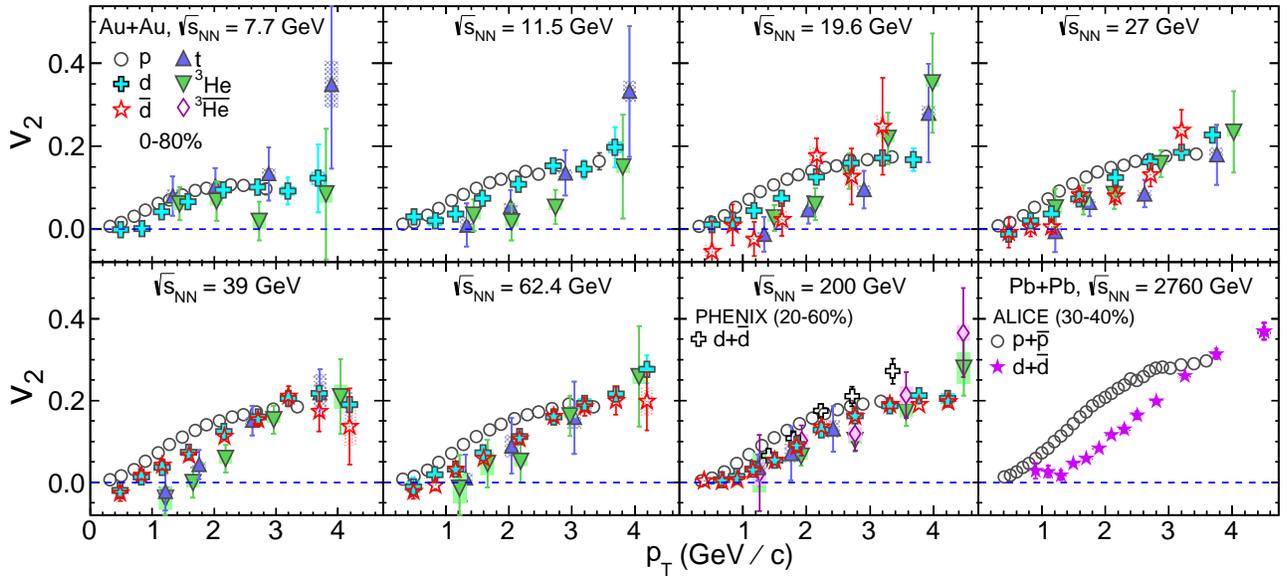}
 \caption{\small {(color online) Mid-rapidity $v_{2}(p_{T})$ for light nuclei
 ($d$, $\overline{d}$, $t$, $^{3}He$, $^{3}\overline{He}$) in
 0-80\%, 20-60\%  and 30-40\% centrality from STAR, PHENIX and ALICE,
 respectively. Proton $v_{2}(p_{T})$ are also shown as open 
 circles~\cite{v2_BES_PID,v2_200_QM,PID_v2_ALICE,PID_v2_PHENIX} 
 for comparison. Lines and boxes at each marker corresponds 
 to statistical and systematic errors, respectively.}}
 \label{fig:v2_pt_minbias}
 \end{figure*}

 \begin{figure*}[t!]
\centering
 \includegraphics[totalheight=7.64cm]{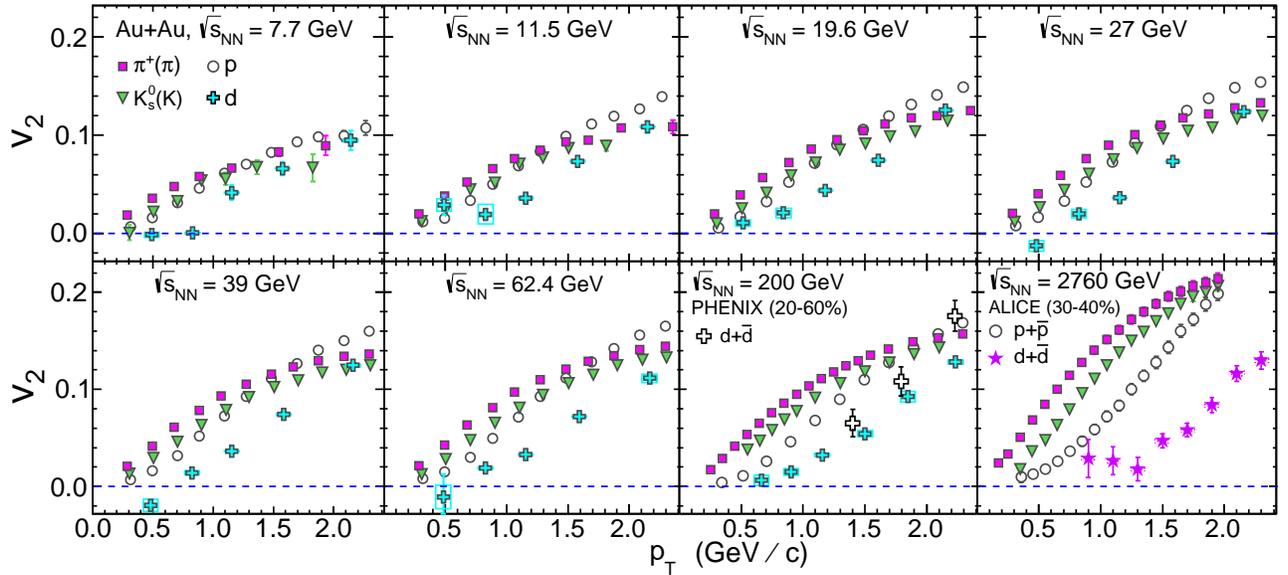}
 \caption{\small {(color online) Mid-rapidity $v_{2}(p_{T})$ for $\pi^{+}$ (squares), 
  $\rm K_{s}^{0}$ (K\ in\ Pb+Pb) (triangles), $p$ (open circles), and
  $d$ (crosses) in 0-80\%, 20-60\%  and 30-40\% centrality from STAR,
  PHENIX and ALICE, respectively.} }
 \label{fig:v2_mass_order}
\end{figure*}

\subsection{Extraction of light nuclei $v_{2}$} 

 In heavy-ion collisions, light nuclei are primarily identified by
 comparing the mean ionization energy loss per unit length ($\langle
 d\rm E/dx \rangle$) in the Time Projection Chamber (TPC) with that 
from the theoretical predictions ($d\rm E/dx|_{theo}$)~\cite{v2_BES_PID,v2_BES_nucl,STAR_TPC,STAR_dAu,
 ALICE_TPC,nucl_ALICE,v2_nucl_ALICE}. Light nuclei are also identified 
 via the time of flight measurement techniques using the Time-of-Flight 
 (TOF) detector~\cite{nucl_PHENIX,STAR_TOF,ALICE_TOF, nucl_ALICE, 
 v2_nucl_ALICE}. 

In the STAR experiment, to identify light nuclei using TPC, a variable
$Z$~\cite{v2_BES_nucl} is defined as 
\begin{equation}
 Z = \log{\large{[} \langle d\rm E/dx\rangle /(d\rm E/dx)|_{theo}\large{]}}.
 \label{eq:defineZ}
\end{equation}
 Then the light nuclei yields are extracted from these
 $Z$-distributions in differential $p_{T}$ and $(\phi-\Psi_{2})$ bins for
 either minimum bias collisions or in selected centrality classes. The
 $(\phi-\Psi_{2})$ distribution is then fitted with a 2$^{nd}$ order
 Fourier function namely,  
\begin{equation}
 \frac{dN}{d(\phi-\Psi_{2})}\ \sim\ 1+2v_{2}\cos{(\phi-\Psi_{2})}.
 \label{eq:Fourier_2nd}
 \end{equation}
 The Fourier co-efficient $v_{2}$ is called elliptic flow and 
is extracted from the fit. As we discussed in the previous
 subsection this measured $v_{2}$ is then corrected with the
 event plane resolution factor ($R_{2}$)~\cite{v2_BES_PID,v2_BES_nucl}. 

 In the ALICE experiment, light nuclei in the low $p_{T}$ region ($<$
 1.0 GeV$/c$ for $d$, $\overline{d}$) are identified by comparing the
 variance ($\sigma_{\langle \frac{d\rm E}{dx} \rangle}$) of the measured
 $\langle d\rm E/dx \rangle$ in the TPC with the corresponding theoretical
 estimate ($d\rm E/dx|_{theo}$)~\cite{nucl_ALICE,v2_nucl_ALICE}. 
 Light nuclei are considered identified if the measured 
 $\langle d\rm E/dx\rangle$ lies within $\pm 3\sigma_{\langle \frac{d\rm E}{dx} \rangle}$ 
of the $d\rm E/dx|_{theo}$.  On the other hand, the light nuclei yield are extracted from the mass squared
 ($m^{2}_{\rm TOF}$) distribution using the TOF detector. The mass of each particle is
 calculated using the time-of-flight ($t$) from the TOF detector and
 the momentum ($\bf{p}$) information from the TPC~\cite{nucl_PHENIX,nucl_ALICE,v2_nucl_ALICE}. Both the ALICE and
 PHENIX experiments use the TOF detector to identify light nuclei at
 high $p_{T}$ ($>$ 1.0 GeV$/c$). The mass of a particle can be
 calculated using the TOF detector as,
\begin{equation}
  m^{2}_{\rm TOF} =   \frac{\bf{p}^{2}}{c^{2}} \big{(} \frac{c^{2}t^{2}}{L^{2}} - 1 \big{)},
 \label{eq:definem2}
\end{equation}
\noindent where the track length $L$ and momentum $\bf{p}$ are
determined with the tracking detectors placed inside magnetic
field~\cite{nucl_ALICE, v2_nucl_ALICE, nucl_PHENIX,PID_v2_PHENIX}. 
After getting the $m^{2}$ for each particle, a selection cut is  
implemented to reject tracks which have their $m^{2}$ several 
$\sigma$ away from the true $m^{2}$ value of the light nuclei, 
as done in the STAR experiment~\cite{v2_BES_nucl}. 
The ALICE experiment, on the other hand defines a quantity $\Delta m$ 
such that, $\Delta m = m_{\rm TOF} - m_{\rm nucl}$, where $m_{\rm nucl}$ 
is the mass of the light nuclei under study. The distribution of 
$\Delta m$ is then fitted with an Gaussian $+$ exponential function 
for signal and an exponential function for the 
background~\cite{v2_nucl_ALICE}. Then $v_{2}$ of light nuclei is 
calculated by fitting the $v_{2}$($\Delta m$) with the weighted 
function,
\begin{equation}
 v_{2}^{\rm Tot}(\Delta m)=v_{2}^{\rm Sig}(\Delta m)\frac{N^{\rm Sig}}{N^{\rm Tot}}(\Delta m)+v_{2}^{\rm Bkg}(\Delta m)
  \frac{N^{\rm Bkg}}{N^{\rm Tot}}(\Delta m),
 \label{eq:v2weight}
\end{equation}
\noindent where the total measured $v_{2}^{\rm Tot}$ is the weighted
 sum of that from the signal ($v_{2}^{\rm Sig}$) and background 
 ($v_{2}^{\rm Bkg}$). The $v_{2}^{\rm Tot}$ of the candidate particles
 are calculated using the scalar product method and corrected for
 the event plane resolution~\cite{v2_nucl_ALICE}.

The PHENIX experiment calculates charged average $v_{2}$ of
(anti)-deuterons as,  
\begin{equation}
v_{2} =  \langle\cos(2(\phi-\Psi_{2})) \rangle / R_{2}
\label{eq:v2calPHENIX}
\end{equation}
The quantity $R_{2}$=$\langle\cos(2(\Psi_{2}-\Psi_{r})) \rangle$ can
readily be identified as the resolution of the event plane
angle~\cite{nucl_PHENIX}. The resolution of full event plane
$\Psi_{2}$ is calculated with sub-event planes ($\Psi^{a}_{2}$,
$\Psi^{b}_{2}$) estimated using two Beam Beam Counter (BBC)
detectors~\cite{PID_v2_PHENIX,nucl_PHENIX}.  
The detailed procedure of calculating the full event plane resolution 
from sub-events are already menioned in the previous subsection.
The large $\eta$ gap between the central TOF and the BBCs 
($\Delta \eta>$ 2.75) reduces the effects of non-flow
significantly~\cite{PID_v2_PHENIX,nucl_PHENIX}. The nuclei $v_{2}$
calculated in PHENIX is also corrected for the  contribution coming
from backgrounds, mainly consisting of mis-identification of other
particles ($e.g.$, protons) as nuclei. A $p_{T}$ dependent correction
factor was applied on the total $v_{2}$ (referred as $v_{2}^{\rm
 Sig+Bkg}(p_{T})$) such that, 
\begin{equation}
 v_{2}^{d(\overline{ d})}(p_{T})=\big{[} v_{2}^{\rm Sig+Bkg}(p_{T}) - (1-R)v_{2}^{\rm Bkg}(p_{T}) \big{]} / R,
 \label{eq:v2PHENIX}
\end{equation}
where $v_{2}^{\rm Sig+Bkg}(p_{T})$ is the measured $v_{2}$ for
$d(\overline{d}) +$ background at a given $p_{T}$,
$v_{2}^{d(\overline{d})}$ is the corrected $v_{2}$ of
$d(\overline{d})$ and $R$ is the ratio of Signal and Signal+Background.

\section{Results and discussion} 

\subsection{General aspects of light nuclei $v_{2}$}
 Figure~\ref{fig:v2_pt_minbias} shows the energy dependence of light 
 \mbox{(anti-)nuclei} $v_{2}$ for $\sqrt{s_{NN}}$ = 7.7, 11.5, 19.6,
 27, 39, 62.4, 200 and 2760 GeV. The panels are arranged by increasing
 energy from left to right and top to bottom. The $p_{T}$
 dependence of $v_{2}$ of $d$, $\overline{d}$, $t$, $^{3}He$
 and $^{3}\overline{He}$ is shown for 0-80\% centrality in STAR,
 20-60\% centrality in PHENIX and 30-40\% centrality in ALICE. Since
 PHENIX and ALICE do not have measurements in the minimum bias
 collisions, we only show the results for mid-central
 collisions. The data points of PHENIX and ALICE correspond to
 inclusive $d$+$\overline{d}$ $v_{2}$.  The general trend of nuclei
 $v_{2}$ of all species is the same$-$ it increases with increasing
 $p_{T}$. The slight difference of $v_{2}$ between STAR and PHENIX is
 due to the difference in centrality ranges. The centrality range for
 PHENIX is 20-60\% and that for STAR is 0-80\%. 
 
 From the trend in Fig.~\ref{fig:v2_pt_minbias} it seems that light 
 nuclei $v_{2}$ shows mass ordering, i.e. heavier particles have
 smaller $v_{2}$ value compared to lighter ones, similar to $v_{2}$ of  identified 
 particles~\cite{v2_BES_PID,PID_v2_ALICE,PID_v2_PHENIX}. In order to see 
 the mass ordering effect more clearly, we restrict the $x-$axis 
 range to 2.5 GeV/c and compare the $v_{2}$ of $d$ with the $v_{2}$ of 
 identified particles such as $\pi^{+}$, $\rm K_{s}^{0}$ (K\ in\
 Pb+Pb) and $p$ as shown in Fig.~\ref{fig:v2_mass_order}. We see that 
 $d$ $v_{2}$ at all collision energies is lower than the $v_{2}$ of the 
 identified hadrons at a fixed value of $p_{T}$. Although mass ordering is a theoretical expectation 
 from the hydrodynamical approach to heavy-ion collisions~\cite{hydro}, 
 coalescence formalism for light nuclei can also give rise to this effect. 
 Recent studies using AMPT and VISHNU hybrid  model suggest that mass 
 ordering is also expected from transport approach to heavy-ion 
 collisions~\cite{ampt_order,urqmd_order}. The $v_{2}$ of light nuclei 
 is negative for some collision energies as shown in
 Fig.~\ref{fig:v2_pt_minbias}. This negative $v_{2}$ is expected to be the outcome of strong 
 radial flow in heavy-ion collisions~\cite{art2}. 

 \begin{figure}[t]
\centering
 \includegraphics[totalheight=6.0cm]{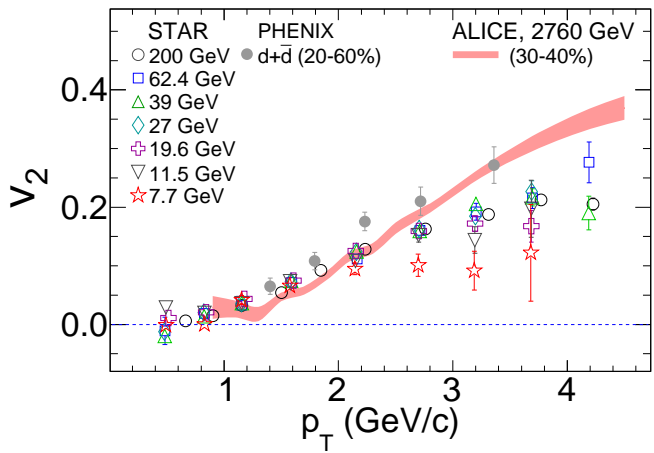}
 \caption{\small {(color online) Energy dependence of mid-rapidity
     $v_{2}(p_{T})$ of $d$ for minimum bias (30-40\% for ALICE) collisions.} }
 \label{fig:v2_deut_enrgy}
\end{figure}

 \begin{figure}[t]
\centering
 \includegraphics[totalheight=6.0cm]{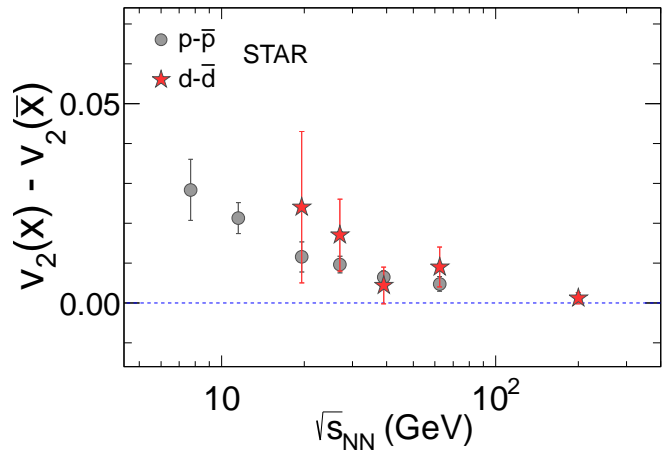}
 \caption{\small {(color online) Difference of $d$ and $\overline{d}$
 $v_{2}(p_{T})$ as a function of collision energy for minimum bias
 Au+Au collisions in STAR.} } 
 \label{fig:v2_ddbar_diff}
\end{figure}

 In order to study the energy dependence of light
 nuclei $v_{2}$, we compare the deuteron $v_{2}$ from
 collision energy $\sqrt{s_{NN}}$ = 7.7 GeV to 2760 GeV as shown in Fig.~\ref{fig:v2_deut_enrgy}. 
 The deuteron $v_{2}$($p_{T}$) shows energy dependence  
 prominently for high $p_{T}$ ($p_{T}>$ 2.4 GeV/c) where $v_{2}$ is 
 highest for top collision energy ($\sqrt{s_{NN}}$ = 2760 GeV) and gradually 
 decreases with decreasing collision energy. This energy dependent trend of 
 light nuclei $v_{2}$ is similar to the energy dependence of identified
 hadron $v_{2}$ where $v_{2}$($p_{T}$) also decreases with decreasing 
 collision energy~\cite{v2_BES_PID}. 

 \begin{figure*}[t]
\centering
 \includegraphics[totalheight=7.9cm]{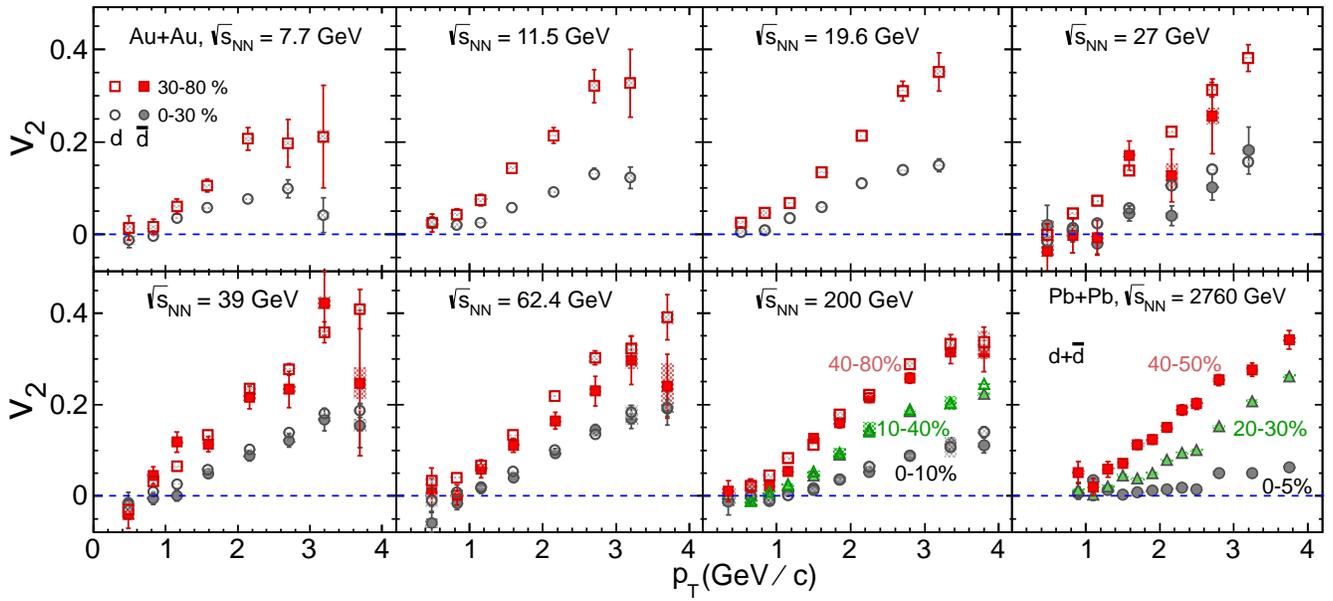}
 \caption{\small {(color online) Centrality dependence of $v_{2}$ of 
 $d(\overline{d})$  as a function of $p_{T}$. }}
 \label{fig:v2_centrality}
\end{figure*}

 The STAR experiment has measured the difference of nuclei ($d$) and
 anti-nuclei ($\overline{d}$) $v_{2}$ for collision energies $\sqrt{s_{NN}}$ 
 = 19.6, 27, 39, 62.4 and 200 GeV~\cite{v2_BES_nucl}. 
 Figure~\ref{fig:v2_ddbar_diff} shows the difference of $d$ and 
 $\overline{d}$ $v_{2}$ as a function of collision energy. For comparison, 
 the difference of proton and anti-proton $v_{2}$ is also shown~\cite{v2_BES_PID}. 
 We observe that the difference of $d$ and $\overline{d}$ $v_{2}$ remains positive 
 for $\sqrt{s_{NN}}$ = 7.7 $-$ 39 GeV. However, for $\sqrt{s_{NN}} \ge$ 62.4 GeV the 
 difference of $d$ and $\overline{d}$ $v_{2}$ is almost zero. The
 difference of $d$ and $\overline{d}$ $v_{2}$ qualitatively follows
 the same trend as seen for difference of $p$ and $\overline{p}$
 $v_{2}$~\cite{v2_BES_PID}. It is easy to infer from simple
 coalescence model that light (anti-)nuclei formed via coalescence of
 (anti-)nucleons, will also reflect similar difference in $v_{2}$ as
 the constituent nucleon and anti-nucleon. The difference in $v_{2}$
 between particles and their antiparticles has been attributed to the
 chiral magnetic effect in finite baryon-density matter~\cite{v2_CMW},
 different $v_{2}$ of produced and transported particles~\cite{v2_Prod},
 different rapidity distributions for quarks and
 antiquarks~\cite{v2_qqbar}, the conservation of baryon number,  
strangeness, and isospin~\cite{v2_conservB}, and different mean-field  
potentials acting on particles and their antiparticles~\cite{v2_meanfield}.

 \begin{figure*}[t]
\centering
 \includegraphics[totalheight=9.7cm]{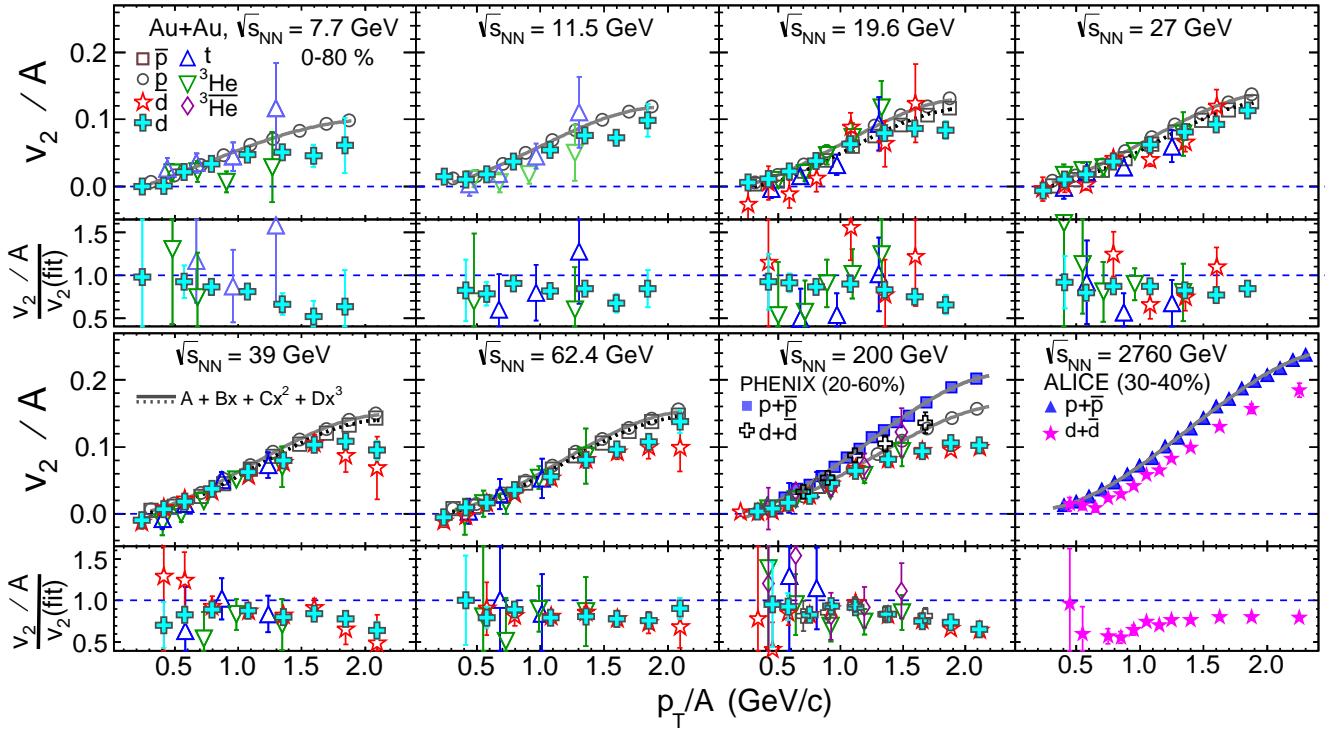}
 \caption{\small {(color online) Atomic mass number scaling $v_{2}/A$
     of light nuclei as a function of $p_{T}/A$ for  STAR (0-80\%), PHENIX
  (20-60\%) and ALICE (30-40\%).} }
 \label{fig:v2_Ascale}
\end{figure*}

The centrality dependence of light nuclei $v_{2}$ measured by the STAR and
ALICE is shown in Fig.~\ref{fig:v2_centrality}. STAR has measured $d$ and $\overline{d}$ $v_{2}$ in two different centrality ranges namely 0-30\% and
 30-80\% for collision energies below $\sqrt{s_{NN}}$ = 200
 GeV. In case of $\sqrt{s_{NN}} =$ 200 GeV, the light nuclei $v_{2}$
 is measured in three different centrality ranges namely 
 0-10\% (central), 10-40\% (mid-central) and 40-80\% (peripheral) as
 high statistics data were available.
 ALICE has measured inclusive $d+\overline{d}$  $v_{2}$ in 6 different
 centrality ranges namely, 0-5\%, 5-10\%, 10-20\%, 20-30\%, 30-40\% and 
 40-50\%. We only present the results from 0-5\%, 20-30\% and 40-50\% 
 centrality from ALICE as shown in Fig.~\ref{fig:v2_centrality}. The
 $v_{2}$ of $d$ shows strong centrality dependence for all collision
 energies studied in the STAR experiment. We observe that more central 
 events has lower $v_{2}$ compared to peripheral
 events. $\overline{d}$ shows the same trend as 
 $d$ for collision energies down to $\sqrt{s_{NN}}$ = 27 GeV. 
 \begin{figure*}[t]
\centering
 \includegraphics[totalheight=9.2cm]{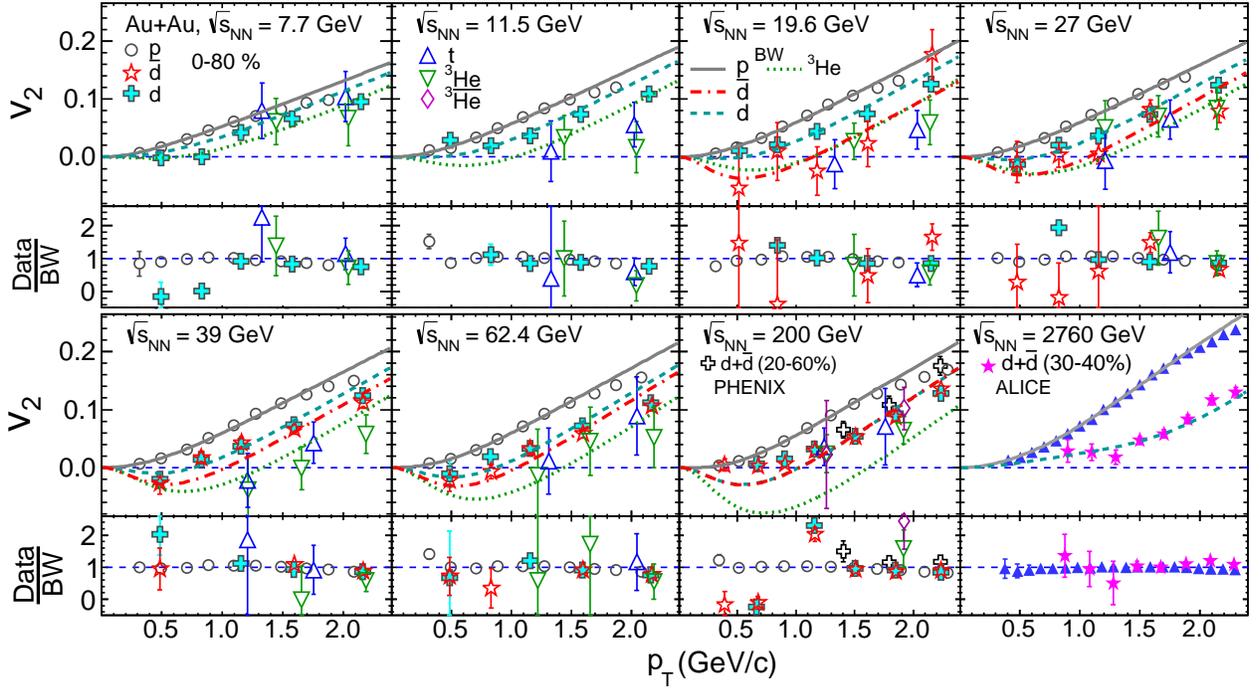}
 \caption{\small {(color online) Light nuclei $v_{2}$ as a function of $p_{T}$
     from blast-wave model (lines). For comparison,
     $p$+$\overline{p}$ $v_{2}$ is also shown. Marker for STAR
     corresponds to 0-80\%, PHENIX corresponds to 20-60\% and ALICE
     corresponds to 30-40\% central events.}}   
 \label{fig:v2_BW}
\end{figure*}

\noindent  The STAR experiment could not study centrality dependence of
 $\overline{d}$  below $\sqrt{s_{NN}}$ = 27 GeV due to limited event
 statistics~\cite{v2_BES_nucl}. Comparing the centrality dependence of
 $d$($\overline{d}$) $v_{2}$ from STAR and ALICE we can see that both
 experiments show strong centrality dependence of light nuclei
 $v_{2}$. The centrality dependence of light nuclei $v_{2}$ is
 analogous to the centrality dependence observed for
 identified nucleon ($p$, $\overline{p}$) $v_{2}$~\cite{PID_cent_STAR,
 PID_cent_ALICE}.

 \begin{figure*}[t]
\centering
 \includegraphics[totalheight=7.2cm]{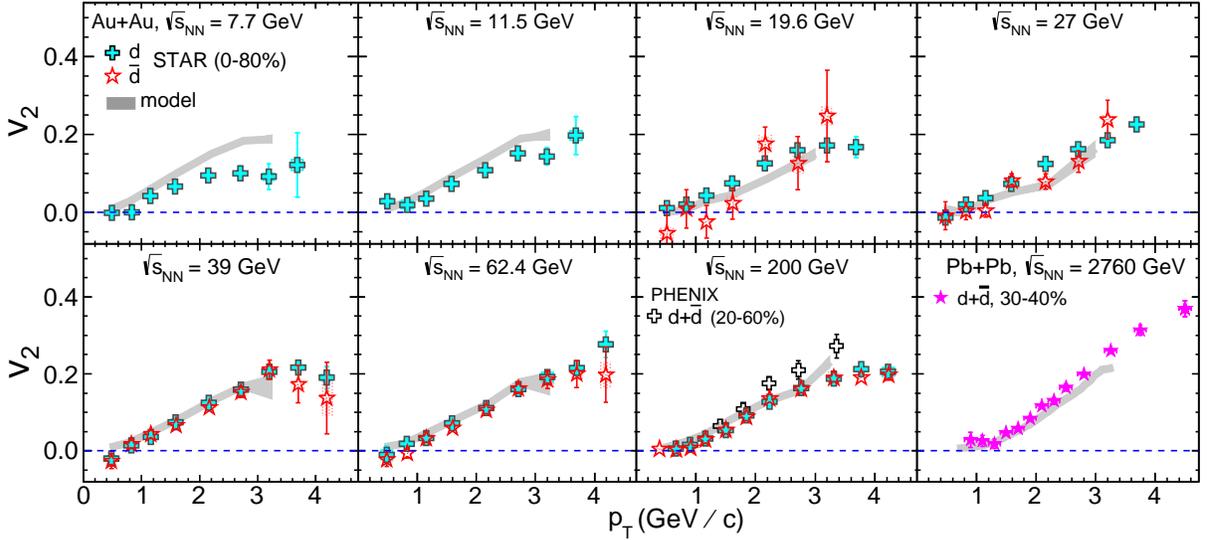}
 \caption{\small {(color online) Light nuclei $v_{2}$ as a function of $p_{T}$
     from AMPT+coalescence model (solid lines). Markers for STAR
     experiment corresponds to 0-80\%, PHENIX corresponds to
     20-60\% and ALICE corresponds to 30-40\% central events.}}
 \label{fig:v2_coalescence}
\end{figure*}

\subsection{Mass number scaling and model comparison}

 \begin{figure*}[t]
\centering
 \includegraphics[totalheight=11.0cm]{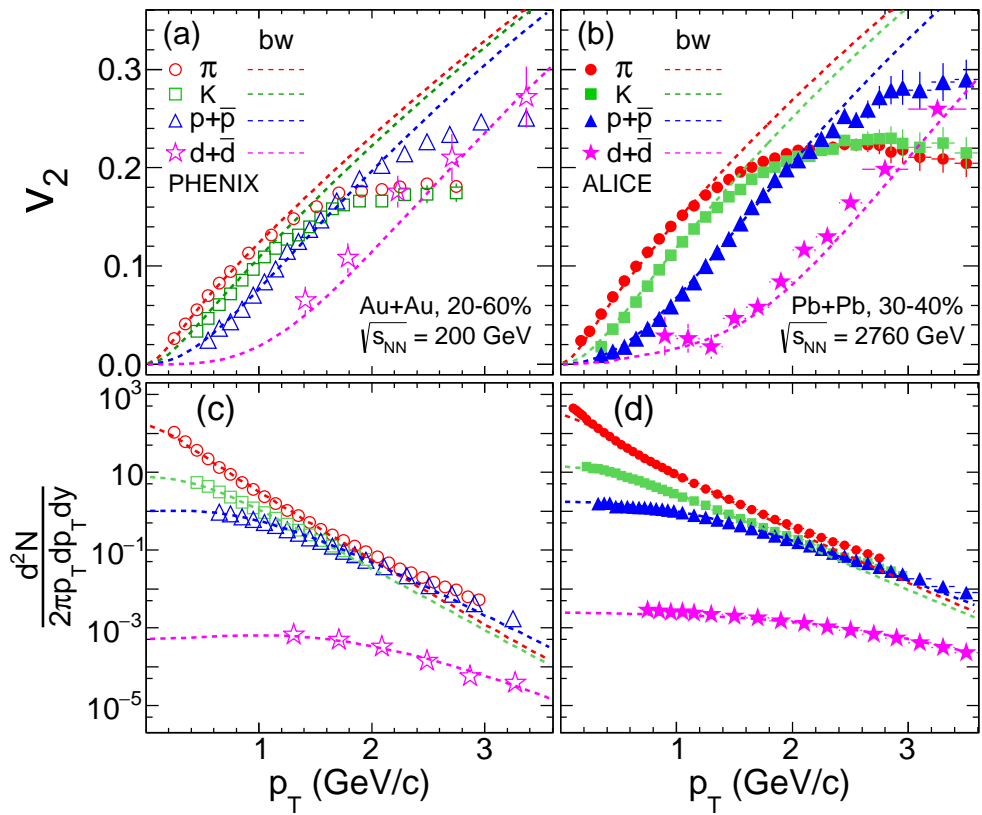}
 \caption{\small {(color online) (a) Blast-wave fit of $\pi$, $K$,
     $p(\overline{p})$, $d(\overline{d})$ $v_{2}$ and (c) $p_{T}$
     spectra from the PHENIX experiment. Same is shown for the ALICE
     experiment in the panel (b) and (d). The $p_{T}$ spectra are
     used from~\cite{PID_pT_PHENIX,PID_pT_ALICE}. Markers for PHENIX
     data corresponds to 20-60\% and markers for ALICE data
     corresponds to 30-40\% central events.}} 
 \label{fig:v2_bw_predict}
\end{figure*}

It is expected from the formulations of coalescence model that if light
nuclei are formed via the coalescence of nucleons then the elliptic
flow of light nuclei, when divided by atomic mass number ($A$), should
scale with the elliptic flow of nucleons ~\cite{firstNPcoal}. 
Therefore, we expect that
the light (anti-)nuclei $v_{2}$ divided by $A$, should scale with
$p$($\overline{p}$) $v_{2}$. Here, we essentially assume that the  $v_{2}$
of \mbox{(anti-)proton} and (anti-)neutron are same as expected from the
observed NCQ scaling of identified particle
$v_{2}$~\cite{v2_BES_PID}. Figure$~\ref{fig:v2_Ascale}$ shows
the atomic mass number scaling of light nuclei $v_{2}$ from STAR,
PHENIX and ALICE experiments. Since ALICE does not have results in 
minimum bias events, so we used both $p$+$\overline{p}$ and 
$d$+$\overline{d}$ $v_{2}$ from 30-40\% centrality range.
We observe that light nuclei $v_{2}$ from STAR and PHENIX show atomic 
mass number scaling up to $p_{T}/A$ $\sim$1.5 GeV/c. 
However, deviation of the scaling of the order of 20\% is
obesrved for $d$+$\overline{d}$ $v_{2}$ from ALICE. The scaling of
light (anit-)nuclei $v_{2}$ with (anti-)proton $v_{2}$ suggests that
light (anit-)nuclei might have formed via coalescence of
(anti-)nucleons at a later stage of the evolution at RHIC energies for 
$p_{T}/A$ up to 1.5 GeV/c~\cite{nucl_prod1, nucl_prod2, nucl_prod3, 
nucl_prod4, nucl_prod5}. However, this simple picture of coalescence
may not be holding for ALICE experiment at LHC energies.
On the contrary, there is another method to produce light
nuclei, for example by thermal production in which it is assumed that 
light nuclei are produced thermally like any other primary
particles~\cite{Ratio_nucl,CFO_nucl}. Various thermal model studies
have successfully reproduced the different ratios of produced
particles as well as light nuclei in heavy-ion
collisions~\cite{Ratio_nucl,CFO_nucl}. 

In order to investigate the success of these models, both STAR and
ALICE has compared the elliptic flow of light nuclei with the
predictions from blast-wave models~\cite{v2_BES_nucl,v2_nucl_ALICE}. 
Figure$~\ref{fig:v2_BW}$ shows the $v_{2}$ of light nuclei predicted 
from blast-wave model using the parameters obtained from fits to the 
identified particles $v_{2}$~\cite{v2_nucl_ALICE,PID_BW_fit}. 
We observe that blast-wave model reproduces $v_{2}$ of light nuclei  
from STAR with moderate success except for low $p_{T}$($<$ 1.0 GeV/c) 
where $v_{2}$ of $d$($\overline{d}$) are under-predicted for all 
collision energies. However, the blast-wave model seems to successfully 
reproduce the $d+\overline{d}$ $v_{2}$ from ALICE. 
The low relative production of light nuclei compared to identified 
nucleons at RHIC collisions energies supports the procedure of light 
nuclei production via coalescence mechanism~\cite{nucl_prod1,nucl_prod2, 
nucl_prod3, nucl_prod4, nucl_prod5}. However, the success of 
blast-wave model in reproducing the nuclei $v_{2}$ at LHC and 
moderate success at RHIC suggest that the light nuclei production is 
also supported by thermal process~\cite{Ratio_nucl,CFO_nucl}. 
The light nuclei production in general might be a more complicated
coalescence process, $e.g.$, coalescence of nucleons in the local 
rest frame of the fluid cell. This scenario might give rise to
deviations from simple $A$ scaling~\cite{v2_BES_nucl}.   

At RHIC energies the light nuclei $v_{2}$ have been compared with results
from a hybrid AMPT+coalescence model~\cite{v2_BES_nucl}. A Multi Phase
Transport (AMPT) model is an event generator with Glauber Monte Carlo
initial state~\cite{AMPT}. The AMPT model includes Zhang's Partonic
Cascade (ZPC) model for initial partonic interactions and A Relativistic
Transport (ART) model for later hadronic interactions~\cite{AMPT}. 
The nucleon phase-space information from the AMPT model is fed to the 
coalescence model to generate light nuclei~\cite{v2_BES_nucl,rihan_thesis}.
Figure~\ref{fig:v2_coalescence} shows the light nuclei $v_{2}$ from
the coalescence model and compared to the data. The coalescence model
prediction for $d$+$\overline{d}$ in Pb+Pb collisions at $\sqrt{s_{NN}}$ 
= 2760 GeV is taken from~\cite{coal_ALICE}. The coalescence model
fairly reproduces the measurement from data for all collision energies
except for the lowest energy $\sqrt{s_{NN}}$ = 7.7 GeV. The AMPT model
generates nucleon $v_{2}$ from both partonic and hadronic interactions
for all the collision energies presented. However, increased hadronic
interactions compared to partonic, at lowest collision energies, is
not implemented in the AMPT$+$coalescence model. This could be the
reason behind the deviation of the data from the model predictions at
lowest collision energy~\cite{v2_BES_PID}. 

We have performed simultaneous fit to the $v_{2}$ and $p_{T}$ spectra 
of identified hadrons + light nuclei using the same blast-wave model 
as used in~\cite{PID_v2_ALICE, v2_nucl_ALICE}. The simultaneous fit of 
$v_{2}$ and $p_{T}$ spectra for measurements from the PHENIX and the 
ALICE experiment are shown in Fig.~\ref{fig:v2_bw_predict}. We find 
that the inclusion of light nuclei results to the fit does not change 
the fit results compared to the blast-wave fit performed only on 
identified hadron $v_{2}$ and $p_{T}$ spectra. This indicates that the 
light nuclei $v_{2}$ and $p_{T}$ spectra is well described by the
blast-wave model.

\subsection{Model prediction of $^{3}$He and $^{4}$He $v_{2}$}

 \begin{figure}[t]
\centering
 \includegraphics[totalheight=12.5cm]{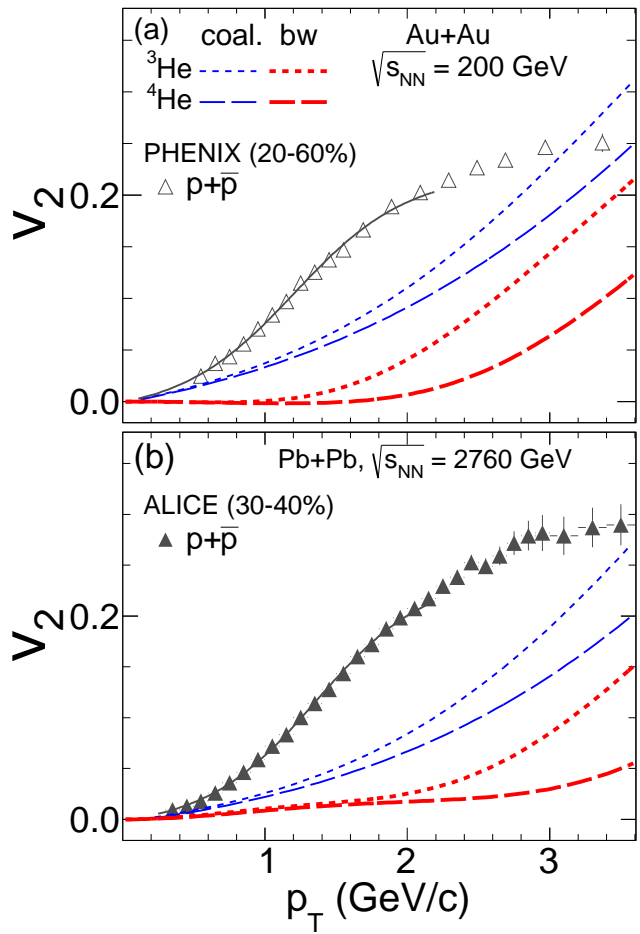}
 \caption{\small {(color online) (a) Coalescence model predictions
     (blue lines) of $^{3}He$ and $^{4}He$ $v_{2}$ for (a) $\sqrt{s_{NN}}$ = 200
     GeV and  (b) for $\sqrt{s_{NN}}$ = 2760 GeV. The blast-wave 
     predictions of $^{3}He$ and $^{4}He$ $v_{2}$ are also shown 
     in red lines.}}
 \label{fig:v2coal_predict}
\end{figure}

We have predicted the $v_{2}$ of $^{3}He$ and $^{4}He$ using the
simple coalescence and blast-wave model. Since protons and neutrons
have similar masses and same number of constituent quarks, they should 
exhibit similar collective behavior and hence, similar magnitude of
$v_{2}$. Therefore, we parametrize the elliptic flow of
$p+\overline{p}$ $v_{2}$ using the fit formula~\cite{v2_fromNCQ}, 
\begin{equation}
  f_{v_{2}(p_{T})}(n) = \frac{a n}{1 + e^{-(p_{T}/n - b)/c} } - dn,
 \label{eq:v2fromNCQ}
\end{equation}
where $a$, $b$, $c$, and $d$ are fit parameters and $n$ is the
constituent quark number of the particle~\cite{v2_fromNCQ}. 
The fit to $p+\overline{p}$ $v_{2}$ (solid lines) from the PHENIX and ALICE
experiment is shown in Fig.~\ref{fig:v2coal_predict}(a) and (b),
respectively. Assuming similar magnitude of neutron
$v_{2}$ as that of proton, we then predict the $v_{2}$ of $^{3}He$ and
$^{4}He$ as,
 \begin{subequations}
 \label{sub:v2d2he4}
  \begin{align}
  v_{2}(p_{T})_{^{3}He}\ \approx\ 3v_{2}(p_{T}/3)_{p}, \\
  v_{2}(p_{T})_{^{4}He}\ \approx\ 4v_{2}(p_{T}/4)_{p}. 
  \end{align}
 \end{subequations}
This simplified coalescence model prediction of $^{3}He$ and $^{4}He$
$v_{2}$ are shown in Fig.~\ref{fig:v2coal_predict}(a) and (b) as blue
(thin-dotted) lines. For comparison, the blast-wave model predicted
$v_{2}$ of $^{3}He$ and $^{4}He$ from the fit parameters obtained 
in Fig.~\ref{fig:v2_bw_predict} are also shown in red (thick-dotted) lines. 
We observe characteristic difference is observed in the prediction of
$^{3}$He and $^{4}$He $v_{2}$ from the coalescence and the blast-wave
model. As one expects from the mass ordering effect of blast-wave
model, the $v_{2}$ of $^{3}He$ and $^{4}He$ almost zero in the
intermediate $p_{T}$ range (1.0 $<p_{T}<$ 2.5 GeV/c). On the other
hand, the simple coalescence model predicts orders of magnitude higher
$v_{2}$ compared to blast-wave for both $^{3}He$ and $^{4}He$ in the
same $p_{T}$ range. Hence, experimental measurements of $^{3}He$ and 
$^{4}He$ $v_{2}$ in future, would significantly improve our knowledge
on the  mechanisms of light nuclei formation in heavy-ion collisions~\cite{nucl_ALICE, STAR_he4_nature, STAR_hypt_science,
  ALICE_hypt_science}.

\section{Summary and conclusions}

 We have presented a review of elliptic flow $v_{2}$ of light nuclei
 ($d$, $t$ and $^{3}He$) and anti-nuclei ($\overline{d}$ and$^{3}\overline{He}$)
 from STAR experiment, and inclusive d+$\overline{d}$ $v_{2}$ from
 PHENIX at RHIC and ALICE at LHC. Similar to identified hadrons, the
 light nuclei $v_{2}$ show a monotonic rise with increasing $p_{T}$ 
 and mass ordering at low $p_{T}$ for all measured collision energies.
 The beam energy dependence of $d$ $v_{2}$ is small at intermediate 
  $p_{T}$ and only prominent at high $p_{T}$, which is similar to the 
  the trend as observed for the charged hadron $v_{2}$. The $v_{2}$ of 
 nuclei and anti-nuclei are of similar magnitude for
 top collision energies at RHIC but at lower collision energies, the
 difference in $v_{2}$ between nuclei and anti-nuclei qualitatively follow
 the difference in proton and anti-proton $v_{2}$. The centrality
 dependence of light \mbox{(anti-)nuclei} $v_{2}(p_{T})$ is similar to 
 that of identified hadrons $v_{2}(p_{T})$. 

 Light (anti-)nuclei $v_{2}$ is found to follow the atomic mass number
 ($A$) scaling for almost all collision energies at RHIC suggesting
 coalescence as the underlying process for the light nuclei production in
 heavy-ion collisions. However, a deviation from mass number scaling at
 the level of 20\% is observed at LHC. 
 This indicates that a simple coalescence process may not be the only
 underlying mechanism for light nuclei production. 
 Furthermore, a transport-plus-coalescence model study is found to
 approximately reproduce the light nuclei $v_{2}$ measured at RHIC and
 LHC. The agreement of coalescence model with the data from PHENIX and
 STAR are imperceptibly better than the blast-wave model. However, at
 the LHC energy, the light nuclei $v_{2}$ is better described by
 blast-wave model rather than the simple coalescence model. The
 coalescence mehcanism, intuitively, should be the prominent process
 of light nuclei production. However, the breaking of mass scaling at
 LHC energy and success of blast-wave model prevent us to draw any
 definitive conclusion on the light nuclei production mechanism.  
   
 We observed orders of magnitude difference in $^{3}He$ and
 $^{4}He$ $v_{2}$ as predicted by blast-wave and coalescence
 model. The blast-wave model predicts almost  
 zero $v_{2}$ for $^{3}He$ and $^{4}He$ up to $p_{T} =$ 2.5 GeV/c,
 whereas the coalescence model predicts significant $v_{2}$ for 
$^{3}He$ and $^{4}He$ at same $p_{T}$ range. Hence, the precise
measurements of $^{3}He$ and $^{4}He$ $v_{2}$ in future can
significantly improve the knowledge of the light nuclei production
mechanism in heavy-ion collisions.

 \section*{Acknowledgements}
We thank STAR collaboration, PHENIX collaboration and ALICE
collaboration for providing the light nuclei $v_{2}$ data and the
model predictions. This work is supported by DAE-BRNS project 
grant No. 2010/21/15-BRNS/2026 and Dr. C. Jena is supported 
by XII$^{th}$ plan project PIC. No. 12-R\&D-NIS-5.11-0300.
The authors declare that there is no conflict of interest
regarding the publication of this paper.

\end{document}